\documentclass[aip,jap,showpacs,twocolumn,reprint,superscriptaddress]{revtex4-1}
\usepackage{amsmath}
\usepackage{amssymb}
\usepackage{graphicx}
\usepackage{hyperref}
\begin{document}
\title{First Principles Study of the Magnetic Properties of LaOMnAs}
\author{Shuai Dong}
\affiliation{Department of Physics, Southeast University, Nanjing 211189, China}
\affiliation{Department of Physics and Astronomy, University of Tennessee, Knoxville, Tennessee 37996, USA}
\affiliation{Materials Science and Technology Division, Oak Ridge National Laboratory, Oak Ridge, Tennessee 37831, USA}
\author{Wei Li}
\affiliation{Shanghai Center for Superconductivity and State Key Laboratory of Functional Materials for Informatics, Institute of Microsystem and Information Technology, Chinese Academy of Sciences, Shanghai 200050, China}
\affiliation{Department of Physics, Fudan University, Shanghai 200433, China}
\author{Xin Huang}
\affiliation{Department of Physics, Southeast University, Nanjing 211189, China}
\author{Elbio Dagotto}
\affiliation{Department of Physics and Astronomy, University of Tennessee, Knoxville, Tennessee 37996, USA}
\affiliation{Materials Science and Technology Division, Oak Ridge National Laboratory, Oak Ridge, Tennessee 37831, USA}
\date{\today}

\begin{abstract}
Recent experiments reported giant magnetoresistance at room temperature in LaOMnAs. Here a density
functional theory calculation is performed to investigate magnetic properties of LaOMnAs.
The ground state is found to be the G-type antiferromagnetic order within the $ab$ plane but coupled
ferromagnetically between planes, in agreement with recent neutron investigations.
The electronic band structures suggest an insulating state which is driven
by the particular G-type magnetic order, while a metallic state accompanies the ferromagnetic order.
This relation between magnetism and conductance may be helpful to qualitatively understand the giant magnetoresistance effects.
\end{abstract}
\maketitle

Since the discovery of high-temperature superconductivity in fluorine-doped LaOFeAs,\cite{Kamihara:Jacs}
several Fe-based pnictide and chalcogenide compounds have been systematically studied.\cite{Johnston:Ap,Stewart:Rmp,Dagotto:Rmp,Dai:Np}
Recently, the Mn-based oxypnictides $R$OMnAs ($R$ is a rare earth, {\it e.g.} La, Nd, Sm) that present
a similar crystal structure as the materials mentioned above have also been synthesized and studied
experimentally.\cite{Shiomi:Prb,Simonson:Prb,Marcinkova:Prb,Kayanuma:Jap,Wildman:Jacs,Emery:Cc,Sun:Epl,Emery:Prb}
In contrast to their Fe-based cousins, at least until now these Mn-based oxypnictides have not shown
any evidence of superconductivity even in doped cases. Instead, giant or even colossal magnetoresistance effects
at room temperature have been observed in these compounds,\cite{Emery:Prb,Emery:Cc,Wildman:Jacs,Sun:Epl}
which reminds us of the well-known phenomenology of the colossal magnetoresistive manganites (Mn-based oxides) with
the perovskite structures.\cite{Dagotto:Prp}

\begin{figure}
\centering	
\includegraphics[width=0.3\textwidth]{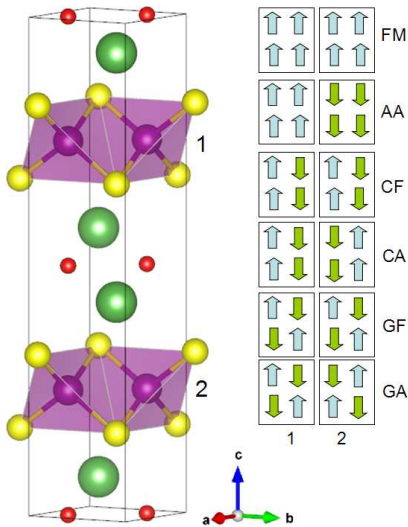}
\caption{(Color online) {\it Left}: crystal structure of a unit cell of
LaOMnAs, which contains four chemical units.
Elements: La (green); O (red); Mn (purple); As (yellow). {\it Right}: sketches of magnetic orders of Mn.
The NN layer indexes are denoted as 1 and 2. The magnetic ground state of $R$OFeAs is CA,
while it is GA or GF for $R$OMnAs.}
\label{crystal}
\end{figure}

Neutron studies have revealed that the magnetic ground state of $R$OMnAs is different from
the isostructural $R$OFeAs.\cite{Marcinkova:Prb,Emery:Prb} %In $R$OFeAs, iron forms a stripe-type spin
%density wave or, in other words, a collinear antiferromagnetic (AFM) state with ferromagnetic (FM) chains
%laying in-plane and coupled antiferromagnetically between them (as in the CA state shown in Fig.~\ref{crystal}).\cite{Cruz:Nat}
%This particular magnetic order can be intuitively understood as induced by the presence of a strong next-nearest-neighbor (NNN) AFM
%exchange $J_2$: $J_2>\frac{|J_1|}{2}$ where $J_1$ is the nearest-neighbor (NN) exchange.\cite{Hu:Prb}
Different from the isostructural LaOFeAs which owns a stripe-like antiferromagnetic (AFM) state (as in the CA one shown in Fig.~\ref{crystal}), $R$OMnAs shows a conventional in-plane G-type AFM order where all NN spins are antiparallel
in-plane.\cite{Emery:Prb}% which requires $|J_2|<\frac{J_1}{2}$ and $J_1$ must be AFM where $J_1$/$J_2$ is the nearest-neighbor (NN) aexchanges.
This magnetic order is also different from magnetic orders in other Fe-based pnictides and chalcogenides, {\it e.g.}
the bi-stripe AFM order found in FeTe\cite{Li:Prb09} and the block-AFM order predicted in KFe$_2$Se$_2$.\cite{Lw:Prb}

To understand the basic physical properties of $R$OMnAs, here we perform a density functional theory (DFT)
calculation to investigate the electronic structures of LaOMnAs, especially its magnetic properties.
Theoretically, LaOMnAs is simpler than NdOMnAs and SmOMnAs within the DFT framework since La$^{3+}$ is
non-magnetic (NM). Fortunately, the most attractive properties of $R$OMnAs, {\it e.g.} its magnetoresistive,
does not rely on the magnetism of the rare-earth component. Thus, the undoped LaOMnAs
provides a good  starting point to study the exotic giant (or colossal) magnetoresistance of $R$OMnAs.
Although Xu \textit{et al}. have already conducted a pioneer DFT calculation on LaO$M$As ($M$=V-Cu),\cite{Xu:Epl}
the real LaOMnAs compound was experimentally studied in detail two years later.\cite{Emery:Cc} Thus,
there are some non-negligible differences between the early DFT predictions and  the experimentally unveiled
facts, such as the lattice constants.% that are $a=4.11398$ \AA{} and $c=9.03044$ \AA{} in experiments
%but $a=4.0355$ \AA{} and $c=8.7822$ \AA{} in the early DFT optimization.\cite{Emery:Cc,Xu:Epl}
%Also, in those early theoretical efforts only the energy differences between the FM state and the simplest
%(staggered) N\'eel AFM states were compared for all $M$'s. Soon it was revealed that the ground state of LaOFeAs
%is actually not this staggered AFM type.\cite{Cruz:Nat}
Later, a DFT calculation by Kayanuma \textit{et al}.
emphasized the band gap of LaOMn$X$ ($X$=P, As, and Sb) but did not focus on the magnetism. Thus, it is
necessary to perform a careful DFT calculation on LaOMnAs with the help of current experimental information
and compare the new results against those previous pioneering theoretical works.

The crystal structure of LaOMnAs is tetragonal ZrCuSiAs-type, with the space group $P4/nmm$,\cite{Emery:Prb}
as shown in Fig.~\ref{crystal}. Each Mn cation is caged by an As tetrahedron. Also each Mn-As layer contains a
square Mn lattice which is effectively isolated from the La-O layer, defining a two-dimensional (2D) network
of Mn cations. %The crystal unit cell, which consists of two Mn layers, contains four chemical units.

The DFT calculation reported here was performed based on the projected augmented wave pseudopotentials
using the Vienna \textit{ab initio} simulation package (VASP).\cite{Blochl:Prb,Kresse:Prb,Kresse:Prb96}
%The valence states include $5d4f6s$, $2s2p$, $3p4s3d$, and $4s4p$ for La, O, Mn, and As, respectively.
The electron-electron interaction is described using the generalized gradient approximation (GGA) method.
The energy cutoff is $500$ eV. The experimental crystal structure (lattice constants $a$-$b$-$c$ and atomic
positions) at $5$ K is adopted.\cite{Emery:Prb} The $\varGamma$-center $k$-mesh is $9\times9\times2$ for
one unit cell, or $6\times6\times2$ for two unit cells (only for the CA and CG cases as shown in Fig.~\ref{crystal}).

To determine the ground state several possible magnetic orders have been tested, as noted in Fig.~\ref{crystal}.
The DFT results are summarized in Table~\ref{dft}. %The FM state is taken as the reference value for the energy.
%All magnetic states considered here have lower energies than the non-magnetic (NM) state, and within the magnetic states
The G-type AFM state has the lowest energy. Here both GA and GF are in-plane G-type AFM for each layer,
but coupled antiferromagnetically and ferromagnetically between NN layers,
respectively.
%\footnote{The GF notation used in the present work is just the so-called C-type antiferromagnetism mentioned in some previous literature.\cite{Sun:Epl}}
The energy difference between GA and GF is of the order of $0.1$ meV,
which is beyond the precision of current DFT calculations. Thus, those states are considered to be degenerate
within our calculation. Similar degenerate characteristics also exist between the CF and CA cases.
Also, a tiny difference occurs between the FM and AA cases.
Thus, from these energy differences, it is safe to conclude that the exchange coupling between NN layers
are very weak ($|J_c|\leqslant1.75$ meV with normalized spins $|S|=1$). To further confirm this point, all atomic positions were relaxed
for the GF and GA cases: in this case the energy difference remained of the order of $0.1$ meV. Our results
agree with the neutron data with regards to the in-plane G-type antiferromagnetism.\cite{Emery:Prb}
The experiments found that LaOMnAs is GF while NdOMnAs is GA possibly due to the influence of the Nd moments,
which also suggested the proximity in energy between those two states.

\begin{table}
\caption{DFT results. Here the FM state is the reference state for the energy.
The energy per Mn is in units of meV. The magnetic moments are in units of $\mu_B$/Mn calculated using Wigner-Seitz spheres as specified by VASP (not very accurate).
The values in brackets for the FM case have been calculated from the total magnetization.
The band gaps are in units of eV.
% The data shown in the last three lines are taken from the previous DFT calculations for comparison.\cite{Xu:Epl}
}
\begin{tabular*}{0.48\textwidth}{@{\extracolsep{\fill}}llllr}
\hline \hline
Magnetic order & Energy & Moment & Band gap\\
\hline
NM & $546.5$ & $0$ & - \\
FM & $0$ & $3.390$($3.559$) & -\\
AA & $3.5$ & $3.338$ & -\\
CF & $-313.5$ & $3.615$ & $0.429$\\
CA & $-313.9$ & $3.614$ & $0.428$\\
GF & $-452.7$ & $3.596$ & $0.524$\\
GA & $-452.6$ & $3.595$ & $0.524$\\
%\hline
%NM & $300$ & $0$ & $-$\\
%FM & $0$ & $2.6$ & not mentioned\\
%GA & $-250$ & $3.1$ & $0.2$\\
\hline \hline
\end{tabular*}
\label{dft}
\end{table}

\begin{figure}
\centering	
\includegraphics[width=0.5\textwidth]{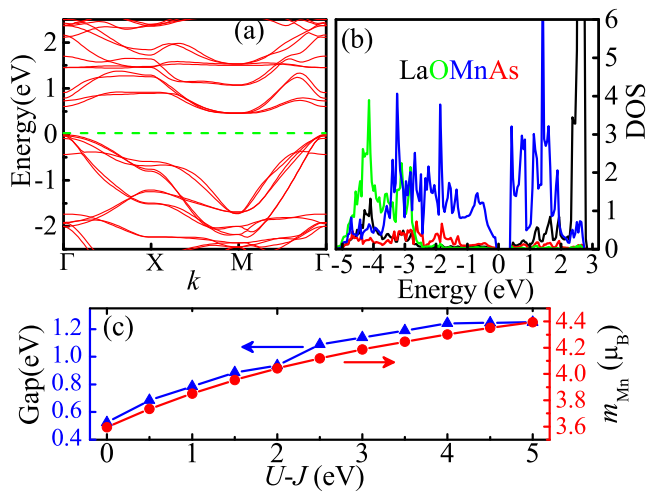}
\caption{(Color online) (a) Band structure near the Fermi level for the
experimentally observed GF state.
(b) The element-resolved DOS. According to (a) and (b), the bands near the Fermi level originate mainly
from the manganese $3d$ orbitals, which hybridizes with the arsenic $4p$ orbitals. (c) The band gap (left axis) and local moment of each Mn (right axis)
as a function of the Hubbard coupling strength acting on the Mn $3d$ electrons. Here $H-J$ is fixed
as $8$ eV for the La $4f$ electrons except for the non-correlated ($U-J=0$) limit.}
\label{band}
\end{figure}

%The local magnetic moment for each Mn is also calculated using Wigner-Seitz spheres as specified in the VASP package,
%which is not very accurate for the plane-wave basis but acceptable.
All calculated magnetic states present strong local moments: about $3.3-3.6$ $\mu_B$ per Mn, which is quite
close to (only a little higher than) the experimental values ($\sim3.34$ $\mu_B$ per Mn at $2$ K).\cite{Emery:Prb}
When the Hubbard $U$ is considered using the GGA+$U$ method, the local moment can further increase to more than $4$ $\mu_B$ per Mn (Fig.~\ref{band}(c)) due to the enhanced localization of $3d$ electrons. According to these results, it seems the GGA method may be even better than the GGA+$U$ one for the LaOMnAs. And these results suggest a high-spin state for Mn due to the strong Hund exchange
although the moments are not ideally large ($5$ $\mu_B$ per Mn).

%The spin-resolved density of states (DOS) have also been calculated, as shown in Fig.~\ref{DOS}. Since
%the couplings between layers are very weak, as stated before, the CA and GA states display almost identical DOS
%as those of  the corresponding CF and GF states, respectively. In addition, the DOS of the AA state is very close
%to the superposition of spin-up and spin-down channels of the FM state. Thus, these three DOS are not shown here.
%The energy gap for GF (or GA) is $0.524$ eV, as shown in Table~\ref{dft}, suggesting semiconducting properties.
%The gaps of the CA and CF states are narrower than in the GA and GF cases. On the other hand, for the AA and FM cases
%the states turn out to be metallic. These results imply that the relationship between magnetism and conductance in LaOMnAs is
%qualitatively analogous as in the case of the well-known double-exchange mechanism of the perovskite manganites.\cite{Dagotto:Prp}

%\begin{figure}
%\centering	
%\includegraphics[width=0.45\textwidth]{Figure2}
%\caption{Spin-resolved DOS for the non-magnetic NM state as well as for various magnetic states, as indicated.
%Here the DOS's for CF and GF are not shown since they are almost identical to the CA and GA cases,
%respectively. The DOS for the AA state is almost identical to the FM one after a superposition
%of the spin-up and down channels.}
%\label{DOS}
%\end{figure}

The band structure of the GA state ({\it i.e.}, the experimentally observed state) is shown in Fig.~\ref{band}(a).
Here the atomic positions have been relaxed in the calculation. The top of the valence bands are
located at the $\varGamma$ point, while the bottom of the conducting bands are at the $M$ point.
Therefore, LaOMnAs is an indirect semiconductor, in agreement with experiments.\cite{Kayanuma:Jap}
Also in experiments, the electronic doping, {\it e.g.} F doping at the oxygen site, will create Fermi surface
pockets around the $M$ point first, while hole doping ({\it e.g.} Ca doping at the La site) will create Fermi
surface pockets around the $\varGamma$ point. This is different from the case of LaOFeAs, which is a bad metal
displaying both electron and hole pockets at the Fermi surface in the undoped case.

The element-resolved DOS further confirms that the bands near the Fermi level are mostly from the Mn's $3d$ orbitals,
which hybridizes with the As's $4p$ orbitals, as shown in Fig.~\ref{band}(b). By contrast, La and O do not contribute
significantly to these bands. Then, in future studies tight-binding models based on multi-$3d$-orbitals as widely
used for Fe-based superconductors can also be used for $R$OMnAs, with only appropriate modifications in the specific values of the
model parameters (such as the crystal-field splitting, hopping amplitudes, electron filling, Hubbard $U$, and Hund coupling).\cite{Dagotto:Rmp}

In the present study, the calculated band gap ($0.524$ eV) is lower than the experimental gap measured
using optical absorption (namely $\sim 1.4$ eV, but this value maybe is not very accurate due to a strong absorption tail)
for LaOMnAs thin films.\cite{Kayanuma:Jap}
%It is well known that frequently DFT calculations without the correlation effect
%incorporated will tend to give smaller gaps than in real materials.
In fact, the earliest DFT calculation
gave an even smaller gap value ($0.2$ eV),\cite{Xu:Epl}. The GGA+$U$ method, incorporating
the effective Hubbard parameter $U-J=1.5$ eV for the Mn $3d$ and $11$ eV for the La $4f$ electrons, gave a larger gap ($0.878$ eV)
but it was still smaller than the experimental one.\cite{Kayanuma:Jap} Here, the GGA+$U$ method is also tested by
tuning the values of $U-J$ to analyze the effect of the Hubbard-type correlation. As shown in Fig.~\ref{band}(c),
the band gap increases with the effective Hubbard parameter $U-J$ for Mn, as expected. For example, the gap is about $1.24$ eV
when $U-J=4$ eV which is already large enough for Mn. In spite of this improvement, the gap width tends to saturation
with further increasing $U-J$ and it is still below the experimental value. Further increase of $U-J$ for the La $4f$ electrons
up to $11$ eV does not alter this result (not shown here). This underestimate of the band gap may be due to the well-known
deficiencies of DFT that tend to favor metallic tendencies, or the inaccuracy of the experimental measurements,
or the difference between thin films and bulk properties.

From the energy difference between the various magnetic states, it is straightforward to extract the superexchange
coefficients by mapping the system to a classical 2D Heisenberg model with the nearest-neighbor exchange $J_1$ and next-nearest-neighbor exchange $J_2$:
%\begin{equation}
%H=J_1\sum_{<i,j>}S_i\cdot S_j+J_2\sum_{[i,k]}S_i\cdot S_k,
%\end{equation}
%where $S$ denotes the normalized spin ($|S|=1$), and $i$, $j$, and $k$ are site indexes; $<>$ and $[]$ stand for the
%NN and NNN neighbors, respectively; $J_1$ and $J_2$ are the NN and NNN exchanges, respectively.
From Table~\ref{dft} and using normalized spins, it is known that $J_1=113.2-114$ meV and $J_2=21.8-22.3$ meV,
both of which are AFM and much stronger than the aforementioned inter-layer coupling.
The NN exchange is the dominant one while the NNN exchange is only about $1/5$
of the NN one, which is different from the case of LaOFeAs where $J_2$/$J_1$ was larger. The physical mechanism of
such strong AFM couplings can be understood by the semi-empirical Goodenough-Kanamori rule considering the half-filled
$3d$ level of the high-spin Mn$^{2+}$ ion.\cite{Goodenough:Jpcs,Kanamori:Jpcs} This strong $J_1$ is responsible
for the high N\'eel temperature $317$ K as well.\cite{Emery:Prb,Emery:Cc}

\begin{figure}
\centering
\includegraphics[width=0.24\textwidth]{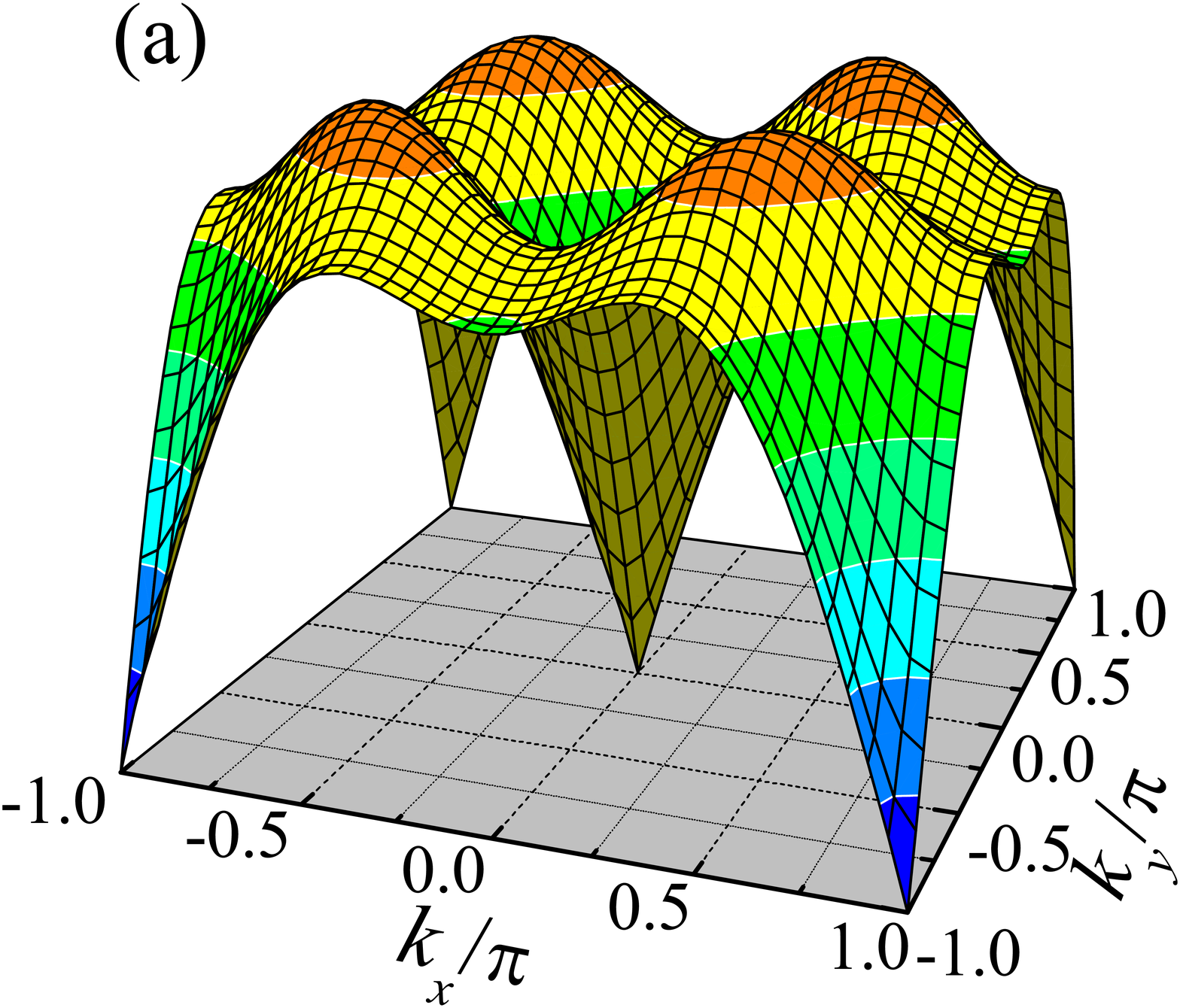}\includegraphics[width=0.24\textwidth]{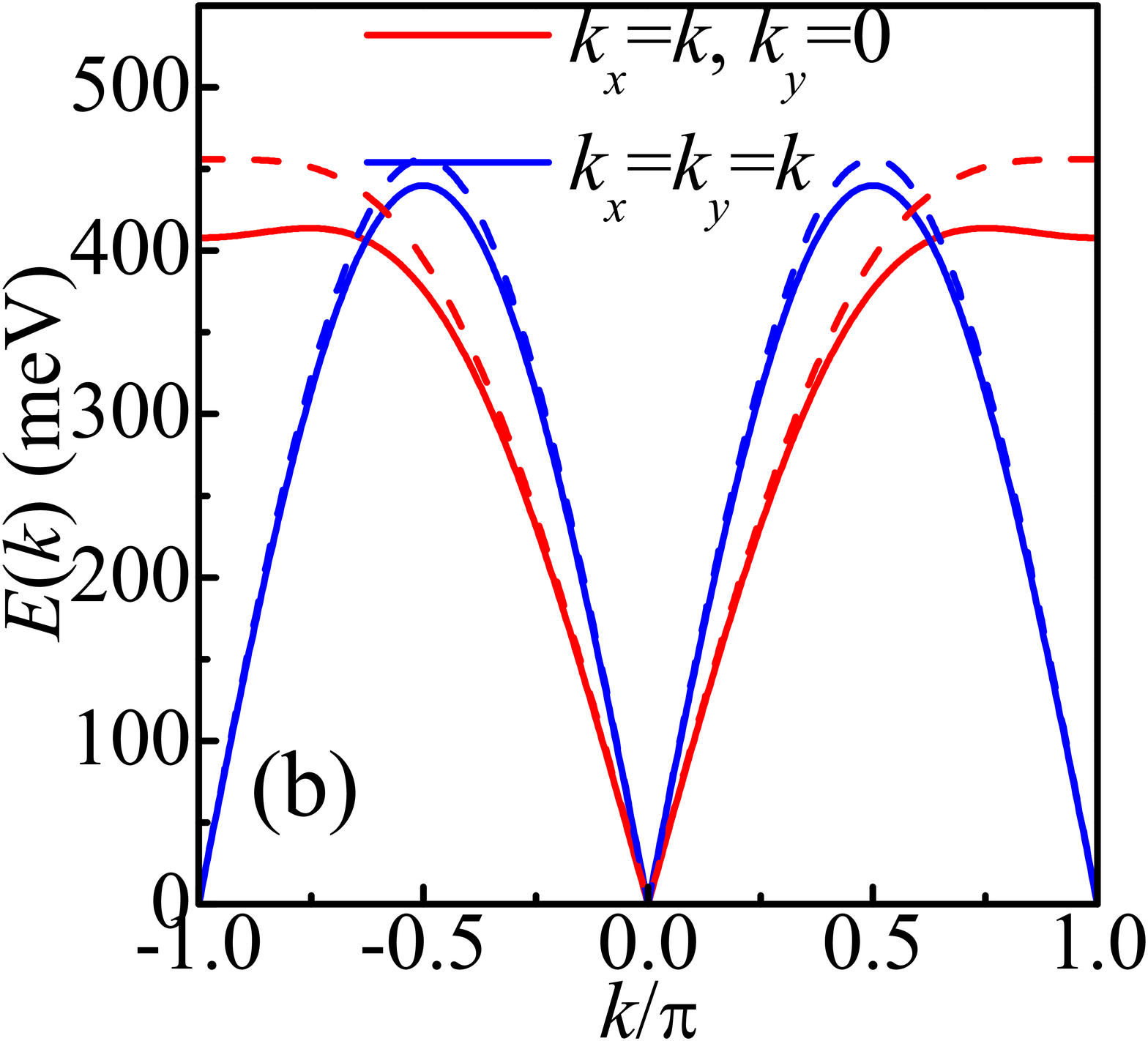}
\caption{(Color online) Spin wave dispersions. (a) In the whole 2D first
Brillouin zone. (b) Along two special lines: (1) $k_y=0$;
(2) $k_x=k_y$. Here the $x-y$ frame is based on the square Mn lattice (the links between NN Mn's).
 The broken curves are dispersions without $J_2$ for comparison.}
\label{sw}
\end{figure}

By using these exchanges, the spin-wave dispersions can be calculated analytically.
%$E=$
%\begin{equation}
%2J_1\sqrt{4[\frac{-J_2}{J_1}(1-\cos{k_x}\cos{k_y})+1]^2-(\cos{k_x}+\cos{k_y})^2},
%\end{equation}
%where $E$ is the magnon energy which depends on the spin wave vector ($k_x$ and $k_y$ are
%the in-plane components). .
As shown in Fig.~\ref{sw}, the spin wave dispersions are typical 2D G-type AFM ones. Due to the NNN exchange $J_2$, the dispersions
become a little softer near the Brillouin boundary when $k_x=0$ or $k_y=0$.

In summary, the oxypnictide LaOMnAs is studied using the first principles calculation. The in-plane magnetic ground
state is confirmed to be the conventional G-type antiferromagnetism, while the coupling between layers is very weak.
The dominant exchange coupling is the nearest-neighbor one while the next-nearest neighbor coupling is relatively
weak. The conductance of LaOMnAs depends on the magnetic order, which may be helpful to understand
the giant magnetoresistance effects observed in these materials.

Work was supported by the 973 Projects of China (2011CB922101) and NSFC (11274060, 51322206).
E.D. was supported by the National Science Foundation Grant No. DMR-1104386.

\end{document}